\documentclass{aastex63}

\let\tablenum\relax
\usepackage{siunitx}
\usepackage{siunitx}
\newcommand\chandra{{\it Chandra}}
\newcommand\ROSAT{{\it ROSAT}}

\newcommand\nustar{{\it NuSTAR}}
\newcommand\nineteenfivetwofive{2MASSi~J0411469+132416}
\newcommand\Mrk{Mrk~504}
\newcommand\FBQS{FBQS~J163302.6+234928}
\newcommand\nineteenfivetwoeight{2MASXi~J0729087+400836}
\newcommand\RBS{RBS~874}
\newcommand\fourC{4C~+50.43}
\newcommand\PG{PG~1302-102}

\def\spose#1{\hbox to 0pt{#1\hss}}
\def\simlt{\mathrel{\spose{\lower 3pt\hbox{$\mathchar"218$}}
     \raise 2.0pt\hbox{$\mathchar"13C$}}}
\def\simgt{\mathrel{\spose{\lower 3pt\hbox{$\mathchar"218$}}
     \raise 2.0pt\hbox{$\mathchar"13E$}}}

\shortauthors{Saade et al.}

\begin{document}

\title{\it{Chandra} \normalfont{Observations of Candidate Sub-Parsec Binary Supermassive Black Holes}}

\correspondingauthor{M. Lynne Saade}
\email{mlsaade@astro.ucla.edu}

\author{M. Lynne Saade}
\affiliation{Department of Physics and Astronomy, University of California, 475 Portola Plaza, Los Angeles, CA 90095}

\author{Daniel Stern}
\affiliation{Jet Propulsion Laboratory, California Institute of Technology, 4800 Oak Grove Drive, Pasadena,
CA 91109}

\author{Murray Brightman}
\affiliation{Cahill Center for Astronomy and Astrophysics, California Institute of Technology, 1216 E. California Blvd., Pasadena, CA 91125}

\author{Zolt\'an Haiman}
\affiliation{Department of Astronomy, Columbia University, New York, NY, 10027}

\author{S.G. Djorgovski}
\affiliation{Cahill Center for Astronomy and Astrophysics, California Institute of Technology, 1216 E. California Blvd., Pasadena, CA 91125}

\author{Daniel D'Orazio}
\affiliation{Department of Astronomy, Harvard University, 60 Garden Street, Cambridge, MA 02138}

\author{K.E.S. Ford}
\affiliation{Department of Astrophysics, American Museum of Natural History, Central Park West at 79th Street, New York, NY 10024}
\affiliation{Department of Science, Borough of Manhattan Community College, City University of New York, New York, NY 10007}
\affiliation{Physics Program, The Graduate Center, City University of New York, New York, NY 10016}

\author{Matthew J. Graham}
\affiliation{Cahill Center for Astronomy and Astrophysics, California Institute of Technology, 1216 E. California Blvd., Pasadena, CA 91125}

\author{Hyunsung D. Jun}
\affiliation{School of Physics, Korea Institute for Advanced Study, 85 Hoegiro, Dongdaemun-gu, Seoul 02455, Korea}

\author{Ralph P. Kraft}
\affiliation{Harvard-Smithsonian Center for Astrophysics, 60 Garden Street, Cambridge, MA 02138}

\author{Barry McKernan} 
\affiliation{Department of Astrophysics, American Museum of Natural History, Central Park West at 79th Street, New York, NY 10024}
\affiliation{Department of Science, Borough of Manhattan Community College, City University of New York, New York, NY 10007}
\affiliation{Physics Program, The Graduate Center, City University of New York, New York, NY 10016}

\author{Alexei Vikhlinin}
\affiliation{Harvard-Smithsonian Center for Astrophysics, 60 Garden Street, Cambridge, MA 02138}

\author{Dominic J. Walton}
\affiliation{Institute of Astronomy, University of Cambridge, Madingley Road, Cambridge CB3 0HA, UK}

\begin{abstract}

We present analysis of \chandra\ X-ray observations of seven quasars that were identified as candidate sub-parsec binary supermassive black hole (SMBH) systems in the Catalina Real-Time Transient Survey (CRTS) based on apparent periodicity in their optical light curves. Simulations predict close-separation accreting SMBH binaries will have different X-ray spectra than single accreting SMBHs, including harder or softer X-ray spectra, ripple-like profiles in the Fe K-$\alpha$ line, and distinct peaks in the spectrum due to the separation of the accretion disk into a circumbinary disk and mini-disks around each SMBH. We obtained \chandra{} observations to test these models and assess whether these quasars could contain binary SMBHs. We instead find that the quasar spectra are all well fit by simple absorbed power law models, with the rest frame 2-10~keV photon indices, $\Gamma$, and the X-ray-to-optical power slopes, $\alpha_{\rm OX}$, indistinguishable from the larger quasar population. This may indicate that these seven quasars are not truly sub-parsec binary SMBH systems, or it may simply reflect that our sample size was too small to robustly detect any differences. Alternatively, the X-ray spectral changes might only be evident at higher energies than probed by {\it Chandra}.  Given the available models and current data, no firm conclusions are drawn. These observations will help motivate and direct further work on theoretical models of binary SMBH systems, such as modeling systems with thinner accretion disks and larger binary separations.
\end{abstract}

\section{Introduction} \label{sec:intro}
Supermassive black holes (SMBHs) are believed to exist in the nuclei of all large galaxies. When galaxies merge, their respective SMBHs generally pair up to form binaries. The SMBH binary separation will then slowly shrink due to dynamical friction and multi-body interactions in asymmetric stellar distributions, as well as due to gas dynamics \citep{berczik2006,mayer2007,gualandris2017}. Multi-body interactions with other SMBHs can also be relevant \citep{ryu2018}. Once the binary reaches sub-parsec scales, the SMBHs can spiral together and merge on timescales shorter than the age of the Universe \citep{begelman1980}. In the final months or years, corresponding to the final 100-1000 orbits before merger, binary SMBHs become strong gravitational wave sources that should be detectable by pulsar timing arrays or future observatories such as the {\it Laser Interferometer Space Antenna} \citep[{\it LISA};][]{amaro-seoane2017}. Close-separation binary SMBHs are therefore important for a range of astrophysical studies, from black hole growth to AGN triggering to gravitational wave physics.  However, few such systems have been conclusively identified to date, and theoretical predictions of their multiwavelength properties vary greatly. In the following, we first review candidate binary SMBHs in the literature (\S~1.1), followed by a review of proposed observational signatures of such systems based on theoretical work (\S~1.2).  We then discuss the motivation behind the work presented here (\S~1.3).

\subsection{Candidate Binary SMBHs in the Literature} \label{subsec:obs}

The first reported candidate sub-parsec binary SMBH was the blazar OJ~287, which displays quasi-periodic optical outbursts every twelve years that could be explained by a binary SMBH slowly decaying under the effects of gravitational radiation \citep{lehto1996}. The binary explanation involves a $1.8 \times 10^{10}\, M_\odot$ primary \citep{valtonen2008} with a $1.4 \times 10^8\, M_\odot$ secondary \citep{valtonen2012} that impacts the primary's accretion disk nearly once per decade. With an optical light curve spanning over 120 years \citep{valtonen2012}, OJ~287 is considered one of the strongest binary SMBH candidates to date.

Additional strong candidate SMBH merger precursors come from systems where multiple, well-separated active galactic nuclei (AGN) are imaged in the same galaxy, such as the two \chandra-detected X-ray AGN in NGC~6420 \citep{komossa2003}, other dual X-ray sources \citep[e.g.,][]{koss2011, koss2012, comerford2015, satyapal2017, pfeifle2019a, pfeifle2019b, hou2019, foord2020}, dual radio sources \citep[e.g.,][]{rodriguez2006, fu2011, fu2015, muller-sanchez2015,kharb2017, rubinur2019}, and dual optical sources \citep[e.g.,][]{liu2011, comerford2012, comerford2013, goulding2019}. For a recent review paper on dual and binary AGNs, see \citet{derosa2020}. As reliably distinguishing two AGN from each other requires the AGN to be well-separated, the dual AGN identified through this method have generally had separations  greater than 1~kpc.  For many years the closest example of a merging SMBH system with multiple AGN detected in imaging was the pair of flat-spectrum radio sources in 0402+379, separated by 7.3~pc \citep{rodriguez2006}.  Recently this has been superceded by a 0.35~pc pair of radio cores in NGC~7674 imaged at 15~GHz using very long baseline interferometry \citep[VLBI;][]{kharb2017}, and it is possible in the near future that subparsec binaries will be resolved with more advanced VLBI like the Event Horizon Telescope \citep[EHT;][]{d'orazio2018a}.

Unusual radio morphologies can also identify candidate merging SMBH systems, since jet precession is a well-established consequence of binary black hole systems \citep[e.g.,][]{gower1982}. VLBI observations have revealed several examples of AGN with jet morphology suggestive of precession, such as S5~1928+738 \citep{kun2014}, 3C~345 \citep{lobanov2005}, and BL~Lacertae \citep{caproni2013}. \citet{tsai2013} and \citet{krause2019} present additional examples based on Australian Telescope Compact Array (ATCA), Very Large Array (VLA), and Multi-Element Radio Linked Interferometer Network (MERLIN) observations. However, these binary SMBH candidates based on jet morphology are considered controversial since Kelvin-Helmholtz instabilities can also mimic the warping of jets due to precession \citep[e.g.,][]{romero1995,lobanov2001}.

More recently, there have been attempts to identify candidate binary SMBHs based on broad emission line profiles, under the assumption that a binary SMBH would modify the lines in a manner similar to a binary star system. \citet{eracleous2012} carried out the first systematic search for quasars with broad line peaks substantially shifted from their nominal wavelengths (by thousands of km s$^{-1}$). They identified 88 such quasars in the Sloan Digital Sky Survey (SDSS), with 14 showing statistically significant changes in H$\beta$ peak velocities. Further candidates from SDSS have been reported based on \ion{Mg}{2} and H$\beta$ emission lines \citep[e.g.,][]{shen2013,ju2013,liu2014,guo2019}. However, it might not be possible to identify highly sub-parsec binaries (e.g., $\leq 0.01$~pc separation) with spectroscopic techniques, as the broad line region (BLR) would be far larger than the orbit of either SMBH, making it potentially insensitive to their movements \citep{shen2010}. Furthermore, several phenomena associated with isolated SMBHs could produce velocity offsets claimed as evidence of binary SMBHs, such as asymmetric reverberation-induced velocity shifts  \citep{barth2015} and unusual BLR geometries \citep{liuJ2016}. \citet{wang2017} updated the analysis of the \citet{ju2013} candidates, as well as observed 1438 more objects with a baseline of 8 years, and found only one candidate with an outlying velocity shift. Subject to these caveats, \citet{li2016} reported a candidate close, centi-parsec (0.018~pc) SMBH binary in the galaxy NGC~5548 based on four decades of spectroscopic monitoring.

Binary SMBH systems can also produce cut-offs or notches in their continuum spectra, evident at rest-frame ultraviolet wavelengths, due to the presence of a secondary black hole truncating or clearing a gap or cavity in the circumbinary disk. \citet{guo2020} recently presented a comprehensive analysis of the spectral energy distributions of $\sim 150$ published candidate periodically variable quasars, but found that the candidate periodic quasars are similar to the control sample matched in redshift and luminosity.

Perhaps the most promising way to identify highly sub-parsec binary SMBHs is through periodicity in AGN light curves. After OJ~287, the first examples of such objects were found in the Catalina Real-Time Transient Survey \citep[CRTS;][]{drake2009}. \citet{graham2015a} reported the first such example, \PG, and \citet{graham2015b} presented a sample of 111 candidate periodic quasars selected from a systematic analysis of 243,500 quasars with well-sampled CRTS light curves\footnote{Note that the newly identified periodic quasars are selected on the basis of sinusoidal variability, which is quite distinct from the regular flaring activity seen in OJ~287.  The sinusoidal light curves are believed due to two SMBHs sharing a single circumbinary accretion disk, which naturally leads to the lower mass black hole having a higher accretion rate and thus the black holes rapidly becoming near equal mass.  OJ~287, on the other hand, has a small secondary SMBH plunging through the accretion disk of a significantly ($\sim 100 \times$) more massive primary twice per orbit \citep{dey2018}.  Therefore, many of the tests for a binary SMBH described below (\S 1.2), and, in particular, tests based on X-ray spectral analysis, are not relevant for OJ~287 despite the significant X-ray observations that exist for this system \citep[e.g.,][]{marscher2011, kushwaha2018, pal2020}.}.  Notably, \PG\, shows the same periodicity in ultraviolet \citep{d'orazio2015a} and mid-IR wavelengths \citep{jun2015b}, as well as in the position angle of its radio jet \citep{qian2018}. \citet{d'orazio2015a} showed that the periodicity of \PG\, could be well explained by a binary SMBH with a mass ratio of $\leq 0.3$ separated by 0.007-0.017~pc (i.e., 1400-3500~AU). Relativistic Doppler boosting and beaming of emission from the secondary SMBH's accretion disk as it orbits a more massive ($\geq 10^{9.1}\, M_\odot$) primary creates the variations observed in the light curve. \citet{liu2018} argued that including data from the All-Automated Sky Survey for Supernovae (ASAS-SN) weakened the case for true periodicity in \PG, though \citet{xin2019} reported on nine additional epochs of simultaneous ultraviolet and optical {\it Swift} observations, finding light curves roughly consistent with the expected trends for the binary model. 

Since the landmark study from \citet{graham2015b}, several other instances of claimed periodicity have been reported in quasar optical light curves, including  examples from the optical photometric databases of PAN-STARRs \citep{liu2019} and the Palomar Transient Factory \citep{charisi2016, dornwallenstein2017}. At higher energies, there have been reports of quasi-periodicity in the gamma-ray light curves of several blazars \citep{sandrinelli2016}, suggestive of SMBH binaries akin to OJ~287, and claims of  modular {\it Swift}-BAT and {\it Swift}-XRT light curves in a local Seyfert galaxy \citep{severgnini2018}. However, concerns with the statistical analyses have been common, including noting the importance of including `red noise' stochastic quasar variability when calculating the false alarm probability \citep{vaughan2016}, as well as proper consideration of the look-elsewhere effect \citep{barth2018}.  While some studies have accounted for these effects in full \citep[e.g.,][]{graham2015b, charisi2016}, the form of the red noise has sometimes been questioned.  For example, \citet{charisi2016} adopted a damped random walk model with Gaussian red noise. M. J. Graham et al.\ (in preparation) presents a more in-depth analysis of these issues, as well as updates the \citet{graham2015b} sample with several more years of photometric monitoring.

\subsection{Predicted Observational Signatures of Binary SMBHs} \label{subsec:theory}

Theoretical models predict a variety of features to be present in accreting binary SMBHs. The two black holes are surrounded by a circumbinary disk, from which accretion streams feed onto mini-disks surrounding each individual black hole. In the interior of the circumbinary disk, a cavity is cleared out by tidal torques from the binary \citep{artymowicz1994}. Figure~8 in \citet{d'ascoli2018} and Figure~3 in \citet{Farris2014} both provide good illustrations of these features. 

Despite the presence of the cavity, the accretion rate onto the SMBHs is not lower than for a single SMBH \citep{d'orazio2013}, and can temporarily exceed that for a single SMBH \citep{rafikov2016}. The total luminosity of the system is also not lower than for a single SMBH \citep{Farris2015b}. For highly unequal mass binaries expected to form as a result of hierarchical galaxy formation \citep[e.g.,][]{volonteri2003},  most of the luminosity arises from accretion onto the secondary black hole \citep{Farris2014, duffell2019}. 

Since accretion is able to proceed in a binary SMBH system as efficiently as with a single SMBH, binary SMBHs should still be able to launch jets. It is well-established that jets will precess under the influence of a black hole binary, creating a jet morphology that is helical on a conical surface \citep{gower1982} or wiggled and knotted \citep{kaastra1992}. The Doppler shift in the synchrotron radiation of the jet will vary periodically due to the precession of the jet, creating periodicity in the radio light curve \citep{kun2014}.

Multiple mechanisms might also cause the accretion onto binary SMBHs to be periodic as well. \citet{d'orazio2013} discuss oscillations in accretion rate created by hydrodynamic interactions of the accretion streams with an overdense lump at the inner edge of the circumbinary disk \citep[see also][]{Farris2014, shi2015}. The resulting periodicity in emission depends on whether emission arises primarily from the circumbinary disk or from the mini-disks --- with implications for the soft vs. hard X-ray light curve as well \citep{tang2018}. Doppler boosts are another possible source of periodic behavior \citep[e.g.,][]{d'orazio2015a}. For a binary where the primary and secondary have equal mass, $M_1 = M_2$, accretion rate oscillations will dominate over Doppler boost oscillations \citep{tang2018}. However, hydrodynamical modulations decline in magnitude with decreasing mass ratio, $q \equiv M_2 / M_1$, such that for $q \leq 0.05$, Doppler effects dominate the periodicity \citep{d'orazio2013, Farris2014, duffell2019}. One last additional source of potential periodic behavior is self-lensing of the accretion flow of one SMBH by the gravitational field of the other \citep{haiman2017, d'orazio2018b}. It is worth distinguishing the relative shapes these modulations introduce onto the light curve, varying from quasi-sinusoidal (Doppler) to bursty (hydrodynamic) to repeating sharp spikes (self-lensing). Current searches for periodic behavior in AGN light curves may only be able to detect the first of these shapes.

In addition to periodicity, potential spectral signatures of binary SMBHs have been considered in the literature. In the UV/optical portion of the electromagnetic spectrum, \citet{roedig2014} argued that the central cavity cleared by the binary would generate a deep notch in the thermal continuum. In contrast, \citet{Farris2015a} and \citet{d'ascoli2018} did not recover the notch, finding that the accretion streams compensate for this gap. \citet{nguyen2016} generated a database of H$\beta$ emission line profiles for sub-parsec binary SMBHs assuming both SMBHs possess mini-disks and illuminate the circumbinary disk. Some of the line profiles generated were highly complex and time-variable, including multiple peaks, but they were also highly dependent on the semi-major axis of the binary as well as the alignment between the mini-disks and the observer. Further work by \citet{nguyen2019} revealed that including the effects of accretion disk winds eliminated the more complex line profiles. This recent work therefore argued that emission line profiles on their own cannot be used to confirm an SMBH binary.

In the X-rays, potential sources of emission in a binary SMBH are the circumbinary disk \citep{tang2018}, hot spots where the accretion streams from the circumbinary disk collide with the mini-disks around each SMBH \citep{roedig2014}, and the mini-disks themselves \citep{Farris2015b}. \citet{roedig2014} predicted that a binary SMBH would have a substantially harder X-ray spectrum than a single SMBH due to thermal emission from the hot spots, with Wien tail emission causing a peak in the spectrum at $\simgt 100$~keV. \citet{ryan2017} found a similar hardening of the X-ray spectrum of binary SMBHs compared to a Novikov-Thorne relativistic thin disk model of an isolated accreting SMBH. However, their hardening takes place at lower energies, at $\simgt 10$~keV. \citet{Farris2015b} predicted that the X-ray spectrum of a binary SMBH will be distinctly harder than a single SMBH that results from a merger. \citet{tang2018} modeled close binaries up until merger, and predicted that the Doppler effect will suppress 1-20 keV emission, while enhancing higher energy emission during the binary phase. They predicted two thermal peaks in the X-ray spectrum, one from the circumbinary disk at $\approx$ 1 keV and another from the mini-disks at $\approx$ 20 keV, with a shallow notch between them. In their simulations of the X-ray spectrum of close binary SMBH systems, \citet{d'ascoli2018} only recovered a single peak at $\approx$ 20 keV, due to Compton reflection by cold, optically-thick matter in the vicinity of the central engine, as commonly seen in isolated SMBH systems \citep[e.g.,][]{george1991}, while emission in the soft X-rays was dominated by the thermal Wien tail from the mini-disks. Their hard X-ray emission was a similar fraction of the total luminosity as that in a single SMBH system, whereas the thermal soft X-ray emission component was more pronounced than in a single SMBH system, though overall they point out their binary SMBH spectrum is more modestly different from a single SMBH spectrum compared to previous models in the literature. Thus, while many results indicate that binary SMBHs should have enhanced X-ray emission relative to single SMBH systems, there is a wide range in the predicted energy at which these enhancements would be seen, ranging from relatively low rest-frame energies of $\sim 1$~keV \citep[e.g.,][]{Farris2015b} to energies more than two orders of magnitude higher \citep[e.g.,][]{roedig2014}. This raises the possibility of binary SMBH systems having dramatically different X-ray spectral indices or ratios of optical to X-ray luminosity than single SMBH systems, which we investigate in \S~3.1 and \S~3.2.

\citet{mckernan2013} also predicted unique patterns of Fe~K-$\alpha$ X-ray spectral lines due to the clearing of the central circumbinary disk by the inward migrating secondary. The broad Fe K-$\alpha$ line profile becomes ripple-shaped due to the presence of an annular gap, with dips in both the red and blue wings of the profile. The energy of these dips depends on the orbital separation. If an inner cavity is cleared in the circumbinary disk, the wings of the broad line profile are suppressed, and if gas piles up at the outer edge of the cavity, double peaks will appear in the broad line profile. \citet{mckernan2013} also modelled the effects of having both the primary and secondary accrete, which creates a secondary broad line component that oscillates on an orbital timescale across the Fe K-$\alpha$ line of the primary. This opens the exciting possibility of spectroscopically detecting very compact binary SMBHs through Doppler shifts in the Fe~K-$\alpha$ lines. 

It should be noted that models in the literature have only considered binaries with close separations less than a few hundred gravitational radii. We discuss the implications of this further in {\S~4.1}.

\subsection{Motivation for this Work} \label{subsec:motiv}

We present the first X-ray spectra for a sample of candidate sub-parsec binary SMBHs. \citet{foord2017} presented a related analysis of a single candidate binary SMBH system {\bf (see \S 4.1)}.  X-rays probe regions of an AGN closer to the central engine than optical and ultraviolet emission, and theoretical work suggests the unique high-energy emission of binary SMBHs could confirm the nature of such systems, as well as provide great promise for probing the compact inner regions where a sub-parsec binary SMBH would be located. However, the wide range of high-energy predictions leaves significant uncertainty on distinguishing features of binary SMBHs. For example, \citet{tang2018} and \citet{d'ascoli2018} have very different predictions for  modeled binaries of similar separation, with the former predicting two peaks in the X-ray spectrum and the latter predicting a single peak. Indeed, no simulation to date has predicted a spectrum with correct and self-consistent thermodynamics, and the results, including the overall scaling of the spectrum (e.g., photon energy, emerging luminosity) is therefore subject to very large uncertainty. This uncertainty reflects the range of different ad-hoc thermodynamical assumptions in different papers.  We therefore crafted our program to provide a first set of observational results to motivate future models, with the goal of searching  for any substantial differences in the spectra of candidate binary SMBHs. Specifically, we sought to observe the maximal number of sources in the minimal amount of observing time. While emission lines like Fe K-$\alpha$ are another promising way to use X-ray spectra to confirm and study close binary SMBH systems \citep{mckernan2013}, the requirements for sufficient detections were beyond the scope of this observational program.

The X-ray data come from a combination of guest observer (GO) and guaranteed time observations (GTO) with the {\it Chandra X-ray Observatory}. The rest of the paper is structured as follows:  \S~2 presents the sample selection and X-ray observations; \S~3 discusses the properties of the sample; \S~4 discusses the results, summarizes our conclusions, and discusses possibilities for future work.  Throughout, we adopt the concordance cosmology, $\Omega_{\rm M} = 0.3$, $\Omega_\Lambda = 0.7$ and $H_0 = 70\, {\rm\,km\,s^{-1}\,Mpc^{-1}}$.

\section{Sample and X-Ray Data} \label{sec:obs}

\subsection{Sample Selection and \chandra\ Observations} \label{subsec:sample}
We selected for observation candidate periodic quasars from \citet{graham2015b}.  In order to ensure high-quality X-ray spectra for this first investigation of a sample of candidate sub-parcec binary SMBH systems in a reasonable observing program (e.g., $\simgt 1000$ counts in $\simlt 10$~ks, per source), we cross-correlated the 111 sources in \citet{graham2015b} with the \ROSAT{} All-Sky Survey Bright Source Catalogue \citep{voges1999}. Using an 18\arcsec\ matching radius (approximately twice the astrometric uncertainty of that survey), nine sources were found to have \ROSAT\ detections.  PG~1302-102, the first candidate sub-pc binary SMBH identified from a periodic optical light curve \citep{graham2015a}, was awarded \chandra\ GTO time in Cycle~18 (P.I. R. Kraft; ObsIDs 19745-19746). Another six sources were awarded GO time in the same cycle (P.I. D. Stern; ObsIDs 19525-19530). Table~\ref{tab:obstable} presents basic properties of the target sample and details of the \chandra\ observations.  The sources range from relatively local (e.g., Mrk~504 at $z =0.036$;\, $\sim 160$~Mpc) to $z > 1$.  \chandra\ exposure times range from 2.7~ks to 15.9~ks. The sources were all observed using the ACIS-S instrument.

\begin{deluxetable*}{lccccccc}[b!]
\tablecaption{Target sample and \chandra\ observation details. \label{tab:obstable}}

\tablecolumns{5}
\tablewidth{0pt}
\tablehead{
\colhead{} &
\colhead{R.A., Dec.} &
\colhead{} & 
\colhead{Obs.} &
\colhead{Exposure} \\
\colhead{Object} &
\colhead{(J2000)} &
\colhead{$z$} &
\colhead{Date} & 
\colhead{(ks)}
}
\startdata
\nineteenfivetwofive{} & 04:11:46.90, +13:24:16.0 & 0.277 & 2017 Mar 26  & 2.7 \\
\nineteenfivetwoeight{} & 07:29:08.71, +40:08:36.6 & 0.074 & 2017 Apr 28 & 7.6 \\
\RBS & 10:30:24.95, +55:16:22.7 & 0.435 & 2017 Sep 4 &15.4 \\
\PG & 13:05:33.01, $-$10:33:19.4 & 0.278 & 2016 Dec 14 & 10.3 \\
\FBQS & 16:33:02.66, +23:49:28.5 & 0.821 & 2017 May 19 & 7.4 \\
\Mrk{} & 17:01:07.76, +29:24:25.0 & 0.036 & 2017 Jun 21 & 5.1 \\
\fourC{} & 17:31:03.60, +50:07:34.0 & 1.070 & 2018 Apr 17 & 15.9 
\enddata

\end{deluxetable*}

\subsection{\chandra\ Data Analysis}

We extracted the \chandra\ spectra using the standard \chandra\ software packages CIAO (version 4.10) with the latest calibration files from CALDB (version 4.8.0).  We extracted spectra using the default spectral grouping carried out by CIAO {\tt specextract}, where the spectrum is grouped with a minimum of 15 counts per bin. The spectra were extracted in circular source regions with 2\arcsec\ radius, with  backgrounds measured in source-free annular regions centered on the targets of inner radius 10\arcsec\ and outer radius 20\arcsec. The spectra were analyzed using XSPEC with the background subtracted and the $\chi^2$ statistic for fitting. We fit each source to a simple power law, adopting the appropriate Galactic absorption from the National Radio Astronomy Observatory \citep[NRAO;][]{dickey1991}. We also investigated including an additional absorption component at the source redshift, but the resulting absorption columns were negligible and the changes in $\chi^2$ did not justify this choice of fit.  The one exception was \nineteenfivetwofive{}. However, the statistical fitting favored physically implausible values $\Gamma\,>\,5$ and $N_{\rm H} \sim 10^{23}\,{\rm cm}^{-2}$, which were particularly unlikely values given the shape of the \nineteenfivetwofive{} X-ray spectrum and the broad Balmer lines in its optical spectrum. Therefore, only Galactic absorption was ultimately included in the final fits for all sources. We did not detect any Fe K-$\alpha$ lines, which is typical for unobscured AGN with with less than $\approx 10^5$ counts \citep{delacalleperez2010}. Table~\ref{tab:xspectra} presents the results of the spectral fitting and Fig.~\ref{fig:allxspecs} shows the \chandra\ X-ray spectra from this study.

\begin{deluxetable*}{lccccc}
\tablecaption{X-ray properties of the sample.\label{tab:xspectra}}
\tablewidth{0pt}
\tablehead{
\colhead{Object} &
\colhead{Counts\tablenotemark{a}}&
\colhead{$\Gamma$\tablenotemark{b}}&
\colhead{$f_{2-10}$\tablenotemark{c}}&
\colhead{$\chi^{2}$/D.O.F.}&
\colhead{$\alpha_{\rm OX}$}
}
\startdata
\nineteenfivetwofive{} & 218  & $2.29^{+0.51}_{-0.47}$   & $0.57^{+0.09}_{-0.12}$ & 7.02/5 & 1.58  \\
\nineteenfivetwoeight{} & 2002  & $1.35^{+0.14}_{-0.14}$  & $3.74^{+0.26}_{-0.24}$ & 53.75/65 & 0.71  \\
\RBS{} & 1891  & $1.53^{+0.12}_{-0.12}$ 
 &$1.01^{+0.05}_{-0.06}$ & 50.76/60 & 1.39 \\
\PG{} & 2379  & $1.59^{+0.13}_{-0.12}$  & $2.00^{+0.12}_{-0.10}$ & 51.56/64 & 1.66 \\
\FBQS{} & 1388 &$1.73^{+0.13}_{-0.12}$  & $1.35^{+0.08}_{-0.07}$ & 57.76/56 & 1.55  \\
\Mrk & 2093  & $1.59^{+0.18}_{-0.17}$  & $3.90^{+0.27}_{-0.34}$ & 57.01/48 & 1.30  \\
\fourC{} & 580  & $1.74^{+0.21}_{-0.20}$  & $0.26^{+0.03}_{-0.03}$ & 20.52/25 & 1.56 \\
\enddata
\tablecomments{D.O.F. stands for `degrees of freedom'. Error bars represent 90\% confidence intervals.}
\tablenotetext{a}{Total counts in the observed 0.5-8~keV band.}
\tablenotetext{b}{Fit in rest-frame 2-10~keV band.}
\tablenotetext{c}{Flux in rest-frame 2-10~keV band, in units of $\rm 10^{-12}\;erg\;cm^{-2}\;s^{-1}$.}
\end{deluxetable*}

\section{Properties of Sample}

\subsection{Black Hole Binary and Active Galaxy Properties}

We recovered black hole masses for each quasar from \citet{graham2015b}, with the exceptions of \Mrk{} and \fourC{} which lacked masses in that paper.  For \Mrk, we use the mass from \citet{ho2008}, while for \fourC, we derived a mass from Palomar optical spectra obtained in September 2019 using the relations in \citet{jun2015a}. \citet{graham2015b} had a suspiciously low mass listed for \nineteenfivetwoeight{}, $\log (M_{\rm BH} / M_{\odot}) = 5.71$, leading to a suspiciously high Eddington ratio. We instead use the mass from \citet{oh2015} for this source. Assuming these single-epoch SMBH masses represent the total binary mass\footnote{The masses were ultimately calibrated using broad line reverberation relations derived from local Seyfert galaxies. As the extent of a quasar BLR is likely far larger than the sub-parsec SMBH binary separations considered here, taking the measured mass to be the total binary mass is reasonable within the caveat concerning the unclear extent to which the original relations apply to luminous quasars.}, we calculated separations for each binary SMBH assuming circular orbits and a mass ratio $q = M_2 / M_1 = 0.5$ where $M_{1,2}$ are the masses of the two SMBHs. The separations are expressed in terms of the gravitational radius $r_g$ where $r_g = G M / c^{2}$, $G$ is the gravitational constant, $M$ is the total binary mass, and $c$ is the speed of light.

There is a well-established correlation between X-ray spectral index $\Gamma$ and the Eddington ratio of AGN, $\lambda_{\rm Edd} = L_{\rm bol} / {L_{\rm Edd}}$, where $L_{\rm bol}$ is the bolometric luminosity and $L_{\rm Edd}$ is the Eddington luminosity \citep{trakhtenbrot2017}. In addition, AGN variability correlates with Eddington ratio \citep{guo2014,rumbaugh2018}. Sources with lower Eddington ratios have harder X-ray spectra and are more variable. As it may have implications for this work, we estimated the Eddington ratios for the sample. Bolometric luminosities were estimated using the 2-10 keV bolometric correction $\kappa_x = L_{\rm bol} / L_{X}$, where $L_{X}$ is the rest frame 2-10~keV X-ray luminosity. $L_{X}$ was calculated using our measured values of $f_{2-10}$ (listed in Table \ref{tab:xspectra})
and the luminosity distance for each source listed in the NASA/IPAC Extragalactic Database (NED). A value of $\kappa_{x}$ = 23, the median bolometric correction for unobscured AGN \citep{lusso2011}, was used to derive the bolometric luminosity. Finally, the bolometric luminosities were divided by the Eddington luminosity for each object, which was estimated using $L_{\rm Edd}=1.26 \times 10^{38}\, (M_{\rm BH} / M_{\odot})\, {\rm erg}\, {\rm s}^{-1}$. 

The CRTS quasars vary widely in their binary properties, with SMBH masses ranging from $4.9 \times 10^{6} M_{\odot}$ to $7.2 \times 10^{9} M_{\odot}$ and binary separations ranging from 57 $r_{g}$ to more than 8400 $r_{g}$. They generally have typical quasar Eddington ratios of a few tenths, with the exception of \fourC{}, which has an Eddington ratio of $\approx 2$ according to our methodology. However, \fourC{} is radio-loud, and radio-loud quasars are known to have elevated X-ray emission \citep[e.g.,][]{zamorani1981, miller2011}, implying that \fourC{} is unlikely to be super-Eddington.  All the other sources in our sample are radio-quiet, with radio fluxes several orders of magnitude lower than \fourC.

These properties, as well as the observed period from CRTS and galaxy morphology, are listed in Table \ref{tab:binary}.  The morphologies are based on visual inspection of PanSTARRS images, supplemented by SDSS images when available.  Five of the sources appear unresolved in the ground-based imaging, with the AGN outshining the host galaxy.  The two lowest redshift sources show a bright, compact nucleus within a disk-like host galaxy.  The table cites published literature that discusses the morphologies of two of the galaxies, both of which show morphological evidence of recent merger activity.

\begin{deluxetable*}{lccccl}
\tablecaption{SMBH binary properties, based on Table 2 in \citet{graham2015b}.\label{tab:binary}}
\tablewidth{0pt}
\tablecolumns{6}
\tablehead{
\colhead{Object} &
\colhead{$\log M_{\rm BH}$} &
\colhead{$r$} &
\colhead{Period}&
\colhead{} &
\colhead{}\\
\colhead{} &
\colhead{($M_\odot$)} &
\colhead{(${r_g}$)} &
\colhead{(d)} &
\colhead{$L_{\rm bol}/L_{\rm Edd}$} &
\colhead{Morphology}} 
\startdata 
\nineteenfivetwofive{}& 8.16 & 922 & 1851 & 0.18 & unresolved; possible close neighbor\\
\nineteenfivetwoeight{}& 7.74 & 1799 & 1612   & 0.18 & nucleated galaxy \\
\RBS{} & 8.43 & 493 & 1515  & 0.50 & unresolved\\
\PG{} & 8.50 & 516 & 1694 & 0.30 & unresolved; merger features \citep{hong2015} \\
\FBQS{} & 9.86 & 57 & 2040 & 0.12 & unresolved \\
\Mrk{} & 6.69 & 8437 & 1408  & 0.46 & nucleated ring galaxy \citep{buta2017} \\
\fourC{} & 8.18 & 677 & 1975 &  2.22 & unresolved \\
\enddata
\tablecomments{Masses are single-epoch mass measurements based on scaling relations for non-binary AGN and are assumed to be the total binary mass (e.g., $M_1 + M_2$) (see text for details). Separations ($r$) assume mass ratios of $q = M_2 / M_1 = 0.5$.  Morphologies are based on PanSTARRS, supplemented with SDSS when available.}
\end{deluxetable*}

\subsection{X-Ray Spectral Indices}
Several of the theoretical models discussed in \S~1.2 predict that merging SMBH systems should have harder X-ray spectra than isolated accreting SMBHs.  We therefore first analyzed the X-ray spectra to determine if the candidate merging SMBH systems had unusual X-ray spectral indices.  Table~\ref{tab:xspectra} presents the results from our fitting, showing that the spectral indices range from 1.35 to 2.29, with a mean value of $1.68 \pm 0.27$.  Such a mean value is on the low (or hard) side for AGN in general, which typically have $\Gamma \sim\,$1.9, but not inconsistent with their full range of $\sim 1.5-2.0$ \citep[e.g.][]{nandra1994, shemmer2008, brightman2013}.

To make this comparison more quantitative, we used a two-sample Kolmogorov-Smirnov (KS) test to compare the spectral indices of our sample with the spectral indices of accreting SMBHs in the BAT AGN Spectroscopic Survey \citep[BASS;][]{ricci2017}, using their tabulated values of $\Gamma _{0.3-10}$ in Table 15. \PG{} was present in the BASS dataset, so it was discarded from the BASS sample before the KS test was performed. The KS test resulted in a p-value of 0.272, which is too large to reject the null hypothesis that our source spectral indices were drawn from the same distribution as the BASS sample. Cutting the BASS sample to only include unobscured AGN (i.e. with $N_{\rm H}<10^{22}\, {\rm cm}^{-2}$) and AGN with Eddington ratios similar to our sample ($0.1 < L/ {L_{\rm Edd}}<1$) resulted in a p-value of 0.107, not changing the results of the KS test. Further cuts on the BASS control sample based on parameters such as SMBH mass and redshift are inadvisable as our sample is too hetergeneous for this to be warranted (see Table \ref{tab:obstable} and Table \ref{tab:binary}). We therefore conclude that these seven candidate merging SMBH systems have X-ray colors, or spectral indices, typical of isolated quasars, at least over the observed $0.5 - 8$~keV range probed by \chandra. To investigate spectral hardening at higher energies would require observations with a satellite sensitive to to $> 10$~keV photons, such as the {\it Nuclear Spectroscopic Telescope Array} \citep[{\it NuSTAR};][]{harrison2013}.

\subsection{Optical-to-X-Ray Luminosities}

We next investigated whether the candidate merging SMBH systems had unusually strong (or weak) X-ray emission, as suggested by some theoretical models.  For this analysis, we considered the relationship between the X-ray luminosity and UV/optical luminosity, as quantified by the $\alpha_{\rm OX}$ parameter \citep[e.g.,][]{tananbaum1979, just2007, lusso2010}. The relation between the two quantities is typically expressed as a linear correlation log($L_{\rm 2~keV}$)= $\gamma$log($L_{2500\si{\angstrom}}$) + $\beta$ \citep{lusso2016}, or in terms of the parameter $\alpha_{\rm OX}$, defined as

\begin{equation} \alpha_{\rm OX}=-\frac{\log(f_{\rm 2~kev}/f_{2500\si{\angstrom}})}{2.605}.
\end{equation}
The mean value for $\alpha_{\rm OX}$ is typically around 1.5 \citep[e.g.][]{lusso2016}.
 
We recovered the monochromatic rest-frame 2~keV flux density from the XSPEC model. For the rest-frame 2500~\si{\angstrom} flux density, we linearly interpolated between the two nearest effective wavelengths present in public archives of large surveys. For most of the sources, this was the FUV and NUV observations from the {\it GALEX} All-Sky Imaging Survey \citep{bianchi2017}, but for the higher redshift sources \FBQS{} and \fourC{}, we used $u$- and $g$-band data from the Sloan Digital Sky Survey and $g$- and $r$-band data from PanSTARRS, respectively. To correct for Galactic extinction, we used the \citet{schlafly2011} reddening map to determine $E(B-V)$ at the location of each source. For the SDSS and PanSTARRS photometry, we used the NASA/IPAC Extragalactic Database (NED) on-line extinction calculator to determine the implied extinction in the optical bands for the \citet{fitzpatrick1999} mean extinction curve. For the {\it GALEX} bands, we used the same NED extinction calculator to determine the CTIO $V$-band extinction, and then converted this to {\it GALEX} FUV and NUV extinctions using the \citet{fitzpatrick1990} parameterization of the UV extinction curve. 
 
\begin{figure}
\figurenum{2}
\plotone{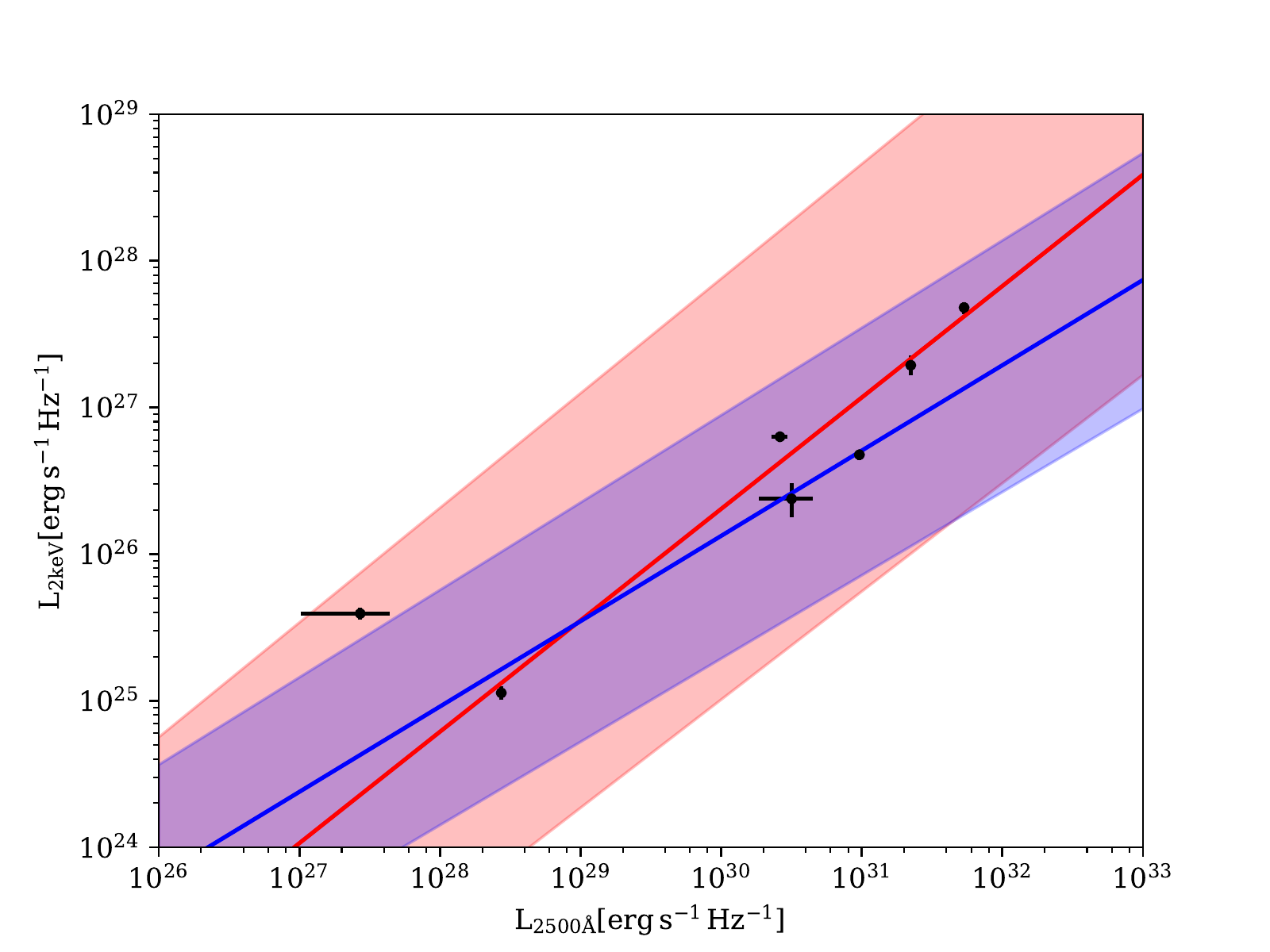}
\caption{$L_{2500\si{\angstrom}}$ vs. $L_{\rm 2keV}$ for our seven sources. Error bars represent propagated 1$\sigma$ uncertainties for optical fluxes, and 90\%\ confidence intervals for X-ray fluxes. In red is the relation from Eq. 6 of \citet{lusso2010} and in blue is the relation for the main sample from Table 2 of \citet{lusso2016}. The shaded regions represent 1$\sigma$ deviations from the mean relation.}
\label{fig:L2500AvsL2keV}
\end{figure}

We then computed the luminosity of each source at 2500~\si{\angstrom} and 2~keV using the luminosity distances listed in NED for each source. Our calculated values of $\alpha_{\rm OX}$ for each source are listed in the final column of Table~\ref{tab:xspectra}. The values range from 0.71 to 1.66, with a mean value of 1.39. Figure~\ref{fig:L2500AvsL2keV} plots  $L_{\rm 2keV}$ as a function of $L_{\rm 2500\si{\angstrom}}$ for our sources, with the corresponding relations from \citet{lusso2010} and \citet{lusso2016}. Out of all the candidate binary SMBH systems investigated here, none are outside the error bars of \citet{lusso2010} and only one (\nineteenfivetwoeight{}) is more than 1$\sigma$ away from the mean trendline of \citet{lusso2016}.  We therefore conclude that these seven candidate merging SMBH systems have optical-to-X-ray luminosities typical of isolated AGN, at least over the observed $0.5 - 8$~keV range probed by \chandra.

\section{Discussion} \label{sec:dis}

We find no obvious differences between the X-ray spectra of the seven sub-parcsec binary quasar candidates in our sample and the X-ray spectra of the quasar population at large, at least over the energies observed by \chandra{}. However, the meaning of this result is unclear. Due to the small sample size studied, unless the differences between binary SMBH and single SMBH spectra are large, there is little chance of detecting a statistically significant difference between binary SMBH quasars and the larger quasar population. We first discuss how details of the binary SMBH models might affect the interpretation of our X-ray results (\S~4.1), and we then discuss potential concerns with the sample itself (\S~4.2).

\subsection{Concerns with Binary SMBH Models} \label{subsec:models}

Assuming the models in the literature are correct, there are several reasons why our sources might truly be sub-parcsec binary SMBH systems but we still would not have expected to observe differences between the {\it Chandra} spectra of our sources and single SMBH systems. The \citet{roedig2014}, \citet{tang2018}, and \citet{d'ascoli2018} models of SMBH binaries have distinct peaks in their X-ray spectra outside the rest-frame 2-10 keV range of energies we observed for most sources, while Figure~3 of \citet{roedig2014} and Figure~17 of \citet{ryan2017}  show only very slight hardening of the X-ray spectra in the 2-10 rest frame keV band. We note that the binary SMBH candidate PSO~J334.2028+01.4075 \citep[identified as a candidate by][]{liu2015} was observed by \citet{foord2017} using \chandra{}, and no peculiarities in the X-ray spectrum were found\footnote{Note, however, that recent work on PSO~J334.2028+01.4075 has weakened the case for periodicity in its optical light curve \citep[e.g.,][]{liu2016} and disfavored its status as a binary SMBH system \citep[e.g.,][]{benke2018}.}. It is therefore possible that the signatures of a binary SMBH lie outside the energy range of \chandra{}.

\citet{Farris2015a} found that the enhancement of emission in binary SMBH systems likely extends to energies lower than the soft X-ray band due to thermal emission from the accretion streams that thread the central cavity. If this emission is enhanced in binary SMBH systems relative to single SMBH systems across the entire optical-UV-X-ray range, then we might not expect our sources to have $\alpha_{\rm OX}$ values distinct from the larger quasar population.

The binary separations, $a$, are also likely an issue for this set of targets being compared to theoretical models. \citet{d'ascoli2018} find that changing the separation of an SMBH binary alters the temperature ratio of the mini-disks relative to the circumbinary disks, with implications for the high-energy spectra. \citet{roedig2014} specifically highlight that their predicted excess thermal X-ray emission would only be clearly distinguishable from ordinary coronal X-rays for binary separations less than 100 gravitational radii, i.e., $< 100\, r_{g}$.  Many of the published models examine very tight sub-parsec binary systems. For example, \citet{d'ascoli2018} modeled a binary with $a = 20\,r_{g}$, while \citet{tang2018} start with a separation $a = 60\, r_{g}$ and evolve it to merger. \citet{roedig2014}, \citet{ryan2017}, and \citet{Farris2015a} considered wider binaries with $a \leq 300\, r_{g}$, $a \approx 100\, r_{g}$, and $a = 100\, r_{g}$, respectively. \citet{Farris2015b} started with a binary at around the same separation as \citet{Farris2015a} and evolved it until merger. 

Considering the binary separations in Table \ref{tab:binary}, only one of our sources has a binary separation $a < 300\, r_g$ (\FBQS{}), and two are separated by more than 1000 $r_g$ (\nineteenfivetwoeight{} and \Mrk{}). Therefore, it is possible that our sources indeed contain binary SMBHs, but with larger separations than what theorists have modeled to date, and at separations where the X-ray spectra are largely indistinguishable from isolated SMBHs.

Furthermore, modeling the extreme regions around isolated accreting SMBHs has many uncertainties, such as the value of the disk viscosity, $\alpha$, in the \citet{Shakura1973} accretion disk model \citep[e.g.,][]{King2007}, or understanding the geometry of the X-ray emitting corona.  Extending these simulations to binary quasars with relativistic velocities is clearly pushing the theoretical models to new, uncertain regimes. Many uncertainties  in models of binary SMBHs arise due to simplifying assumptions. Numerical constraints mean modeled accretion disks are often thicker than AGN disks are in actuality. This shifts the disk Mach number and therefore the frequency of emitted radiation \citep{tang2018}. Effectively, these simulated disks are much hotter than in reality and so the thermal X-ray emission predicted in these simulations needs to be scaled down to lower photon energies.

Similarly, coronae, which are believed responsible for AGN X-ray emission through inverse Compton up-scattering of thermal disk photons, are generally not modeled, but are simply either painted on \citep[e.g.,][]{d'ascoli2018} or modeled as thermal emission \citep[e.g.,][]{Farris2015a, Farris2015b, tang2018}. \citet{d'ascoli2018} note that the manner in which they introduce the corona into their model would lead to an underestimate of hard inverse Compton X-rays and an overestimate of softer thermal X-rays, while \citet{tang2018} note their lack of a true corona would overestimate the ability of the Doppler effect to suppress lower energy X-rays. A first principles description of AGN coronae is not present at this time, and models have difficulty generating observed spectral energy distributions (SEDs) even for single isolated SMBHs. This lack of understanding of the corona introduces complications for direct comparisons of the theoretical SEDs of binary SMBHs to the SED of a single SMBH. Finally, we note that most of the theoretical models use a fiducial SMBH mass of $10^8\, M_\odot$, while our sample ranges from $10^6$ to $10^9\, M_\odot$. This likely has  some implications in terms of the expected slope and peak energy of the disk emission. However, it should be noted the dependence of the disk thermal emission on SMBH mass is not as strong as the dependence on binary separation and disk Mach number \citep[e.g.,][]{Farris2015a,Farris2015b}.

\subsection{The True Nature of the Sources} \label{subsec:sourcenature}

It is uncertain whether or not the seven CRTS quasars in our sample are truly periodic sources. Quasars are known to vary stochastically. The power spectrum of these fluctuations is broad, with the power increasing at low frequencies (``red noise''). The spectrum is often approximated as a power law, $P(f) \propto f^{-\alpha}$, with $\alpha \simgt 1$ over long timescales \citep{vaughan2013}. It is possible for this  noise to generate apparent periodicity when there is none; e.g., \citet{vaughan2016} generated false periodicity in simulated CRTS light curves of \PG{} even though the quasar's output was generated as a damped random walk (DRW) or Gaussian noise. This can occur even for well-sampled data as long as the number of period cycles observed is small \citep{barth2018}. It is also the case when searching for an effect within a wide parameter space where the true location of the effect is unknown, statistically significant detections will happen by pure chance, the so-called ``look-elsewhere effect'' \citep{gross2010}. Properly accounting for the false alarm probability due to the look-elsewhere effect requires a noise model, and if the true stochastic variability contains more power than the best-fit DRW light curves, the purported periodicity can disappear with further observations. 

The original \citet{graham2015b} survey considered CRTS light curves for 243,500 quasars, looking for a strong Keplerian periodic signal with at least 1.5 cycles over a baseline of nine years. Though simulated data sets assuming stochastic variability (e.g., red noise)  produced no equivalent candidates, implying a low likelihood of spurious detections, the short sampling time relative to the best-fit periods raises a natural concern for false positives.  

In addition, even if the periodic behavior observed by CRTS is real, this does not necessarily mean that the quasars in our sample are all sub-parsec binaries. A hotspot on the accretion disk could produce a periodic light curve, with the caveat that many of the mechanisms that might produce a disk hotspot involve an SMBH binary \citep[see][and references therein]{dorazio2015b}. There are alternative explanations for periodicity that involve only a single SMBH.  For example, SMBHs are all expected to have non-zero angular momentum, and so  Lense-Thirring precession will be important if the accretion disk is offset from the equatorial plane of the rotating black hole \citep{bardeen1975}. This could cause both the relativistic jet and the inner accretion disk to precess and create periodic (or quasi-periodic) variability in the optical light curve. Frequent misalignments between the accretion disk and black hole axis are theoretically expected to occur \citep{hopkins2010, hopkins2012}. \citet{graham2015b}, using the results of \citet{ulubay2009} and \citet{tremaine2014}, find the precession period of a warped AGN disk is within an order of magnitude of the potential periods of our sources (assuming an SMBH mass of $10^{8}M_\odot$). However, the precession is damped on a timescale that is short compared to typical AGN lifetimes \citep{tremaine2014, graham2015b}.  Thus, Lense-Thirring precession in an AGN would be rarely observed.

Finally, and perhaps relatedly, the observed quasar periodicity might be caused by the same processes that cause quasi-periodic oscillations (QPOs) in black hole X-ray binaries \citep[BHXBs---i.e., binary systems including a stellar mass black hole; for a recent review of black hole QPOs, see][]{motta2016}. \citet{graham2015b} noted that naively scaling the low-frequency $\sim 1$~Hz QPO of the $\sim 12 M_\odot$ microquasar GRS~1915+105 \citep[see][and references therein]{yan2013} to the estimated mass of \PG{} yields a QPO period that overlaps the observed periodicity of \PG{}'s optical light curve. On the other hand, with the physics of QPOs still uncertain, and the wide range of frequencies spanned by low-frequency QPOs for a given source \citep{wijnands1999}, it is not clear that low-frequency QPOs scale linearly with black hole mass. As one example, \citet{menou2001} note that ionization instabilities, one postulated source of QPOs in BHXBs, will be much more important for stellar mass binary black holes than for SMBHs. In addition, we note that recent work shows that QPOs in BHXBs appear more associated with the inverse Compton X-ray emission from the corona, and not with the thermal accretion disk component \citep[e.g.,][]{remillard2006, ingram2009, ingram2011}. Quasars also have cooler accretion disks than BHXBs, with emission that peaks at rest-frame UV energies rather than the X-ray energies of BHXBs.  Therefore, naively scaling the physics of QPOs from BHXBs to SMBHs might not produce periodic light curves at optical wavelengths.

\subsection{Conclusions} 

We  find  no  obvious  differences  between the  X-ray  spectra of  the  seven  candidate  sub-parcsec binary  SMBHs  in our sample and the X-ray spectra of the quasar population at large, at least over the energies observed by {\it Chandra}. Many theoretical models predict differences in the X-ray spectra of close binary SMBHs, though the models have a wide range of predictions, and the models are not all consistent with each other.  Furthermore, most of the models investigate binaries with closer separations than we estimate for our sample.  This implies inconclusive results from our survey:  the observed sample may or may not indeed be sub-parsec binary SMBH systems.

For future work, analyses at other wavebands might be useful, such as monitoring the UV/optical spectra of candidate binary SMBHs for kinematic variability and periodicity, and/or searching for the proposed UV/optical ``notch'' in the spectral energy distribution due to the inner gap in the circumbinary accretion disk \citep{roedig2014}.  Note, however, that more recent simulations have failed to recover that gap \citep[e.g.,][]{Farris2015b, d'ascoli2018}.  Several models predict spectral hardening of binary SMBHs might lie at higher energies than the bands investigated by \chandra, so further investigations of candidate binary SMBH sources with \nustar{} is an enticing option. However, as some predicted signatures lie as far out as 100 keV \citep{roedig2014}, it is unclear whether even \nustar{} will be able to detect them.

\acknowledgements
We thank the anonymous referee for a very thorough report, which has improved the paper. Support for this work was provided by the National Aeronautics and Space Administration (NASA) through {\it Chandra} Award Number 18700580 issued by the {\it Chandra} X-ray Center, which is operated by the Smithsonian Astrophysical Observatory for and on behalf of the NASA under contract NAS8-03060. The work of DS was carried out at the Jet Propulsion Laboratory, California Institute of Technology, under a contract with NASA. ZH acknowledges support from NSF grant 1715661 and NASA grants 80NSSC19K0149 and NNX17AL82G. MJG and SGD were supported in part by the NASA grant 16-ADAP16-0232, and the NSF grants AST-1413600, AST-1518308, and AST-1749235. DJD acknowledges funding from the Institute for Theory and Computation Fellowship. KESF and BM were supported by NSF grant 1831412. HDJ  was supported by the Basic Science Research Program through the National Research Foundation of Korea, funded by the Ministry of Education (NRF-2017R1A6A3A04005158).

\appendix 

\section{X-Ray Spectra of Sample}

\begin{figure*}
\figurenum{2}

\gridline{\fig{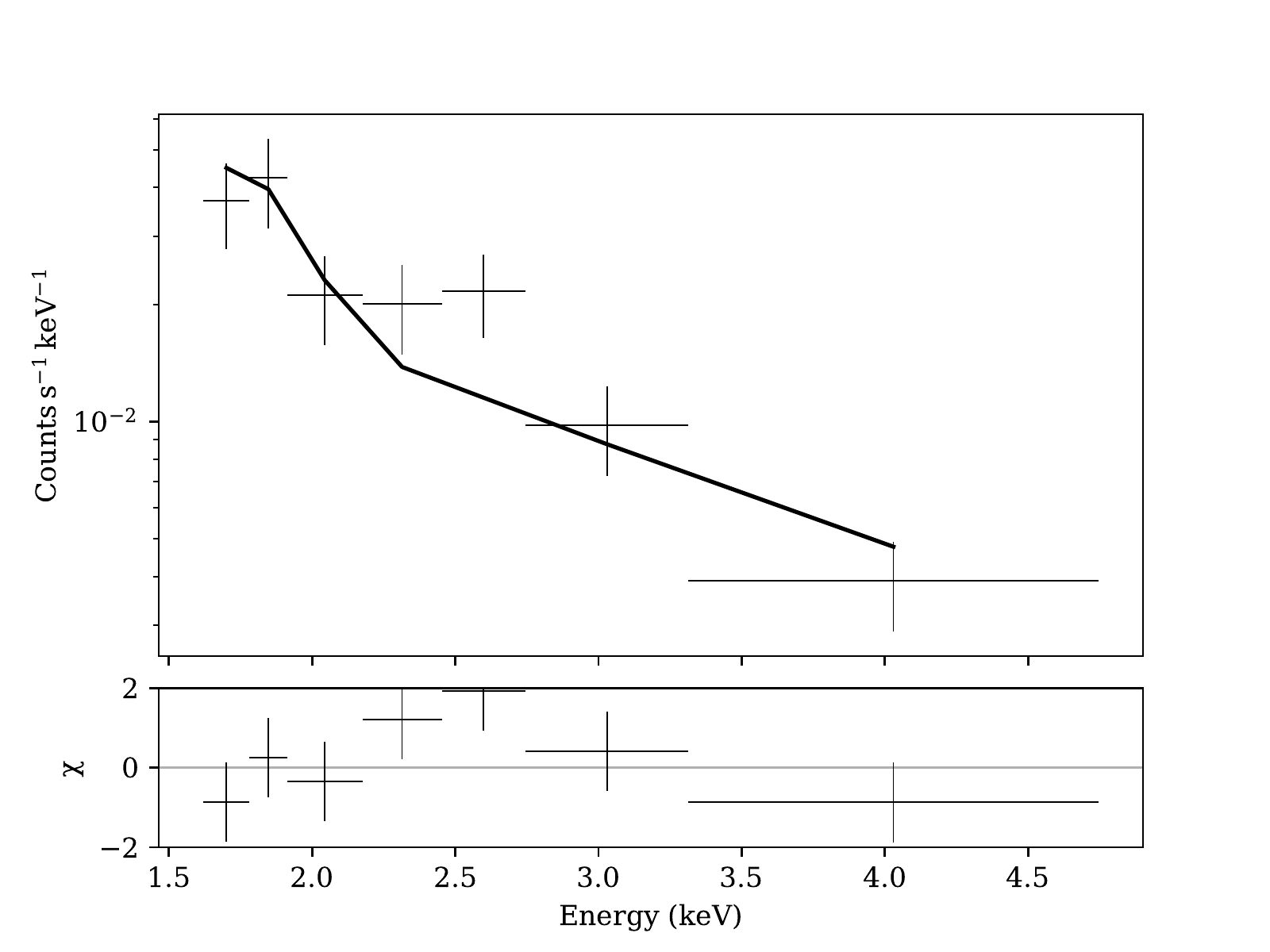}{0.3\textwidth}{(a)}
         \fig{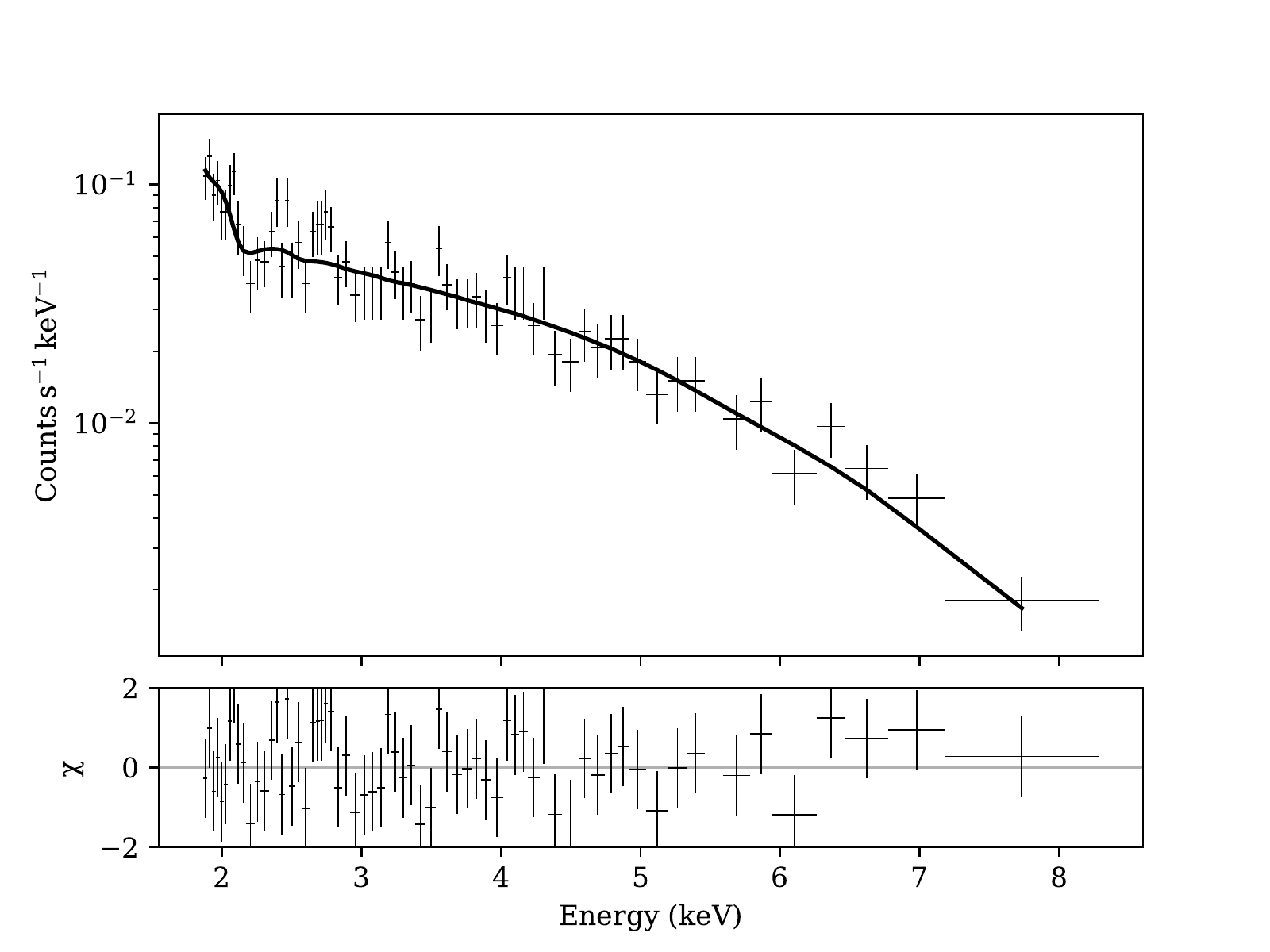}{0.3\textwidth}{(b)}
         \fig{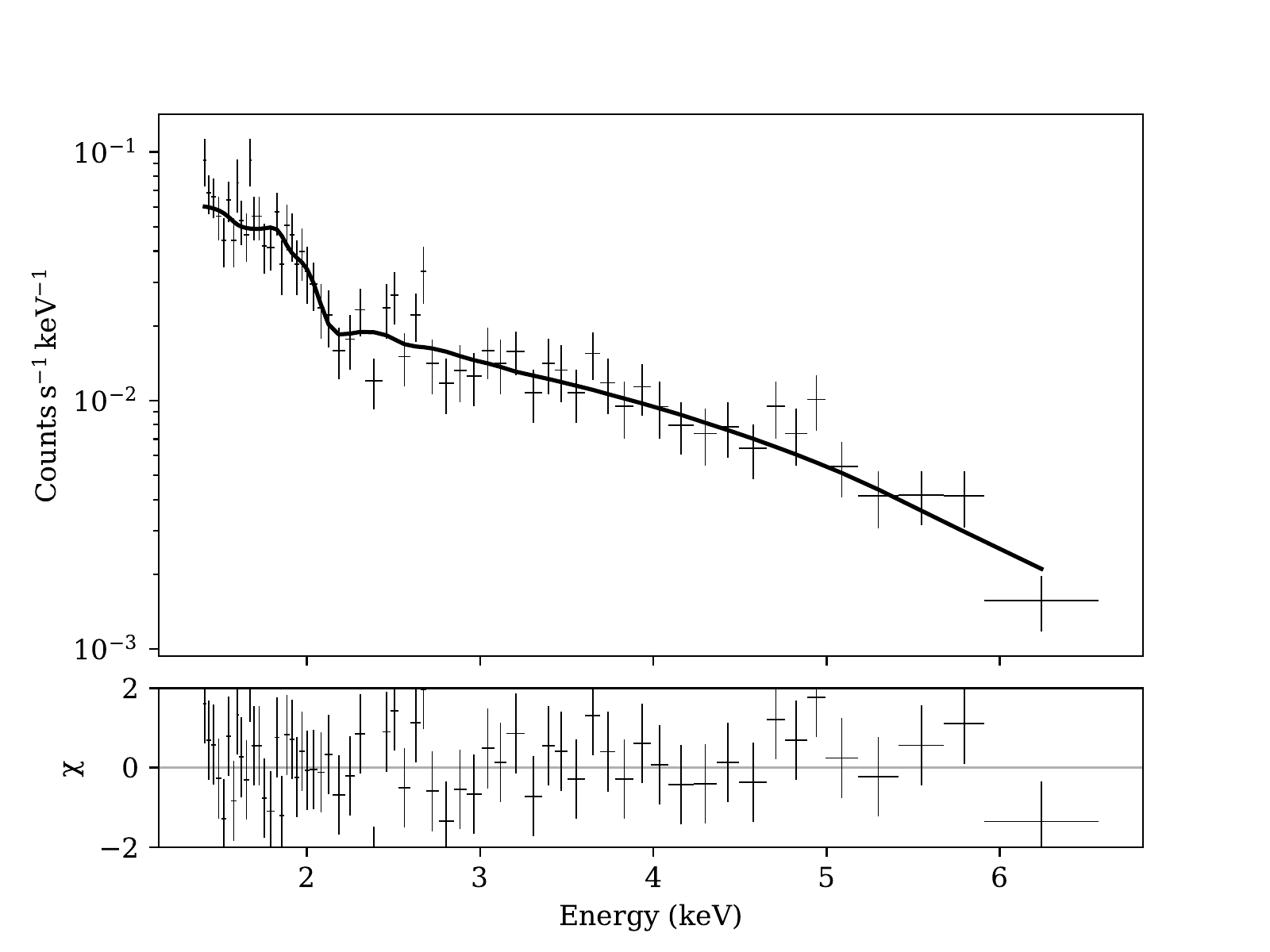}{0.3\textwidth}{(c)}
         }
\gridline{\fig{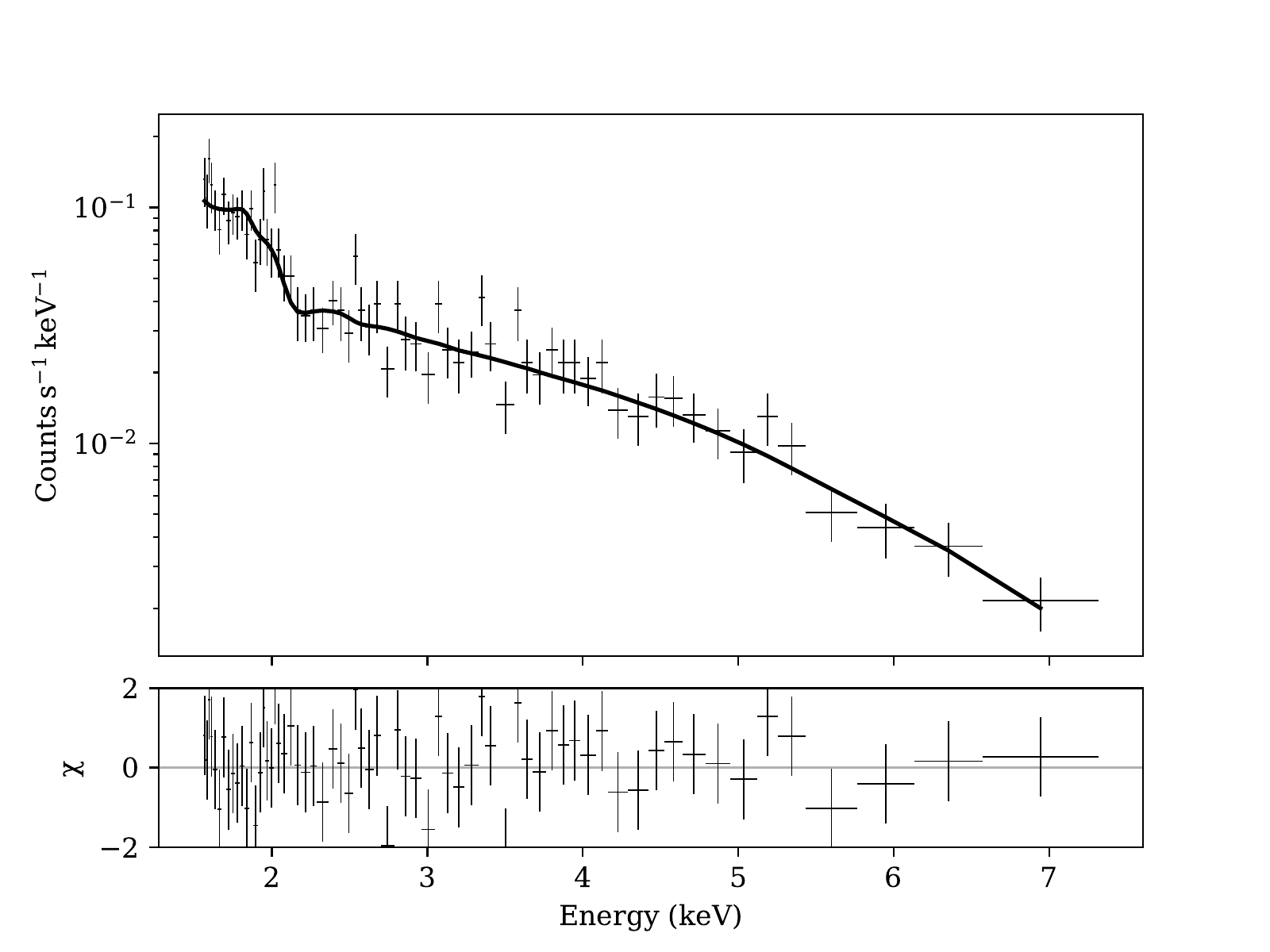}{0.3\textwidth}{(d)}
          \fig{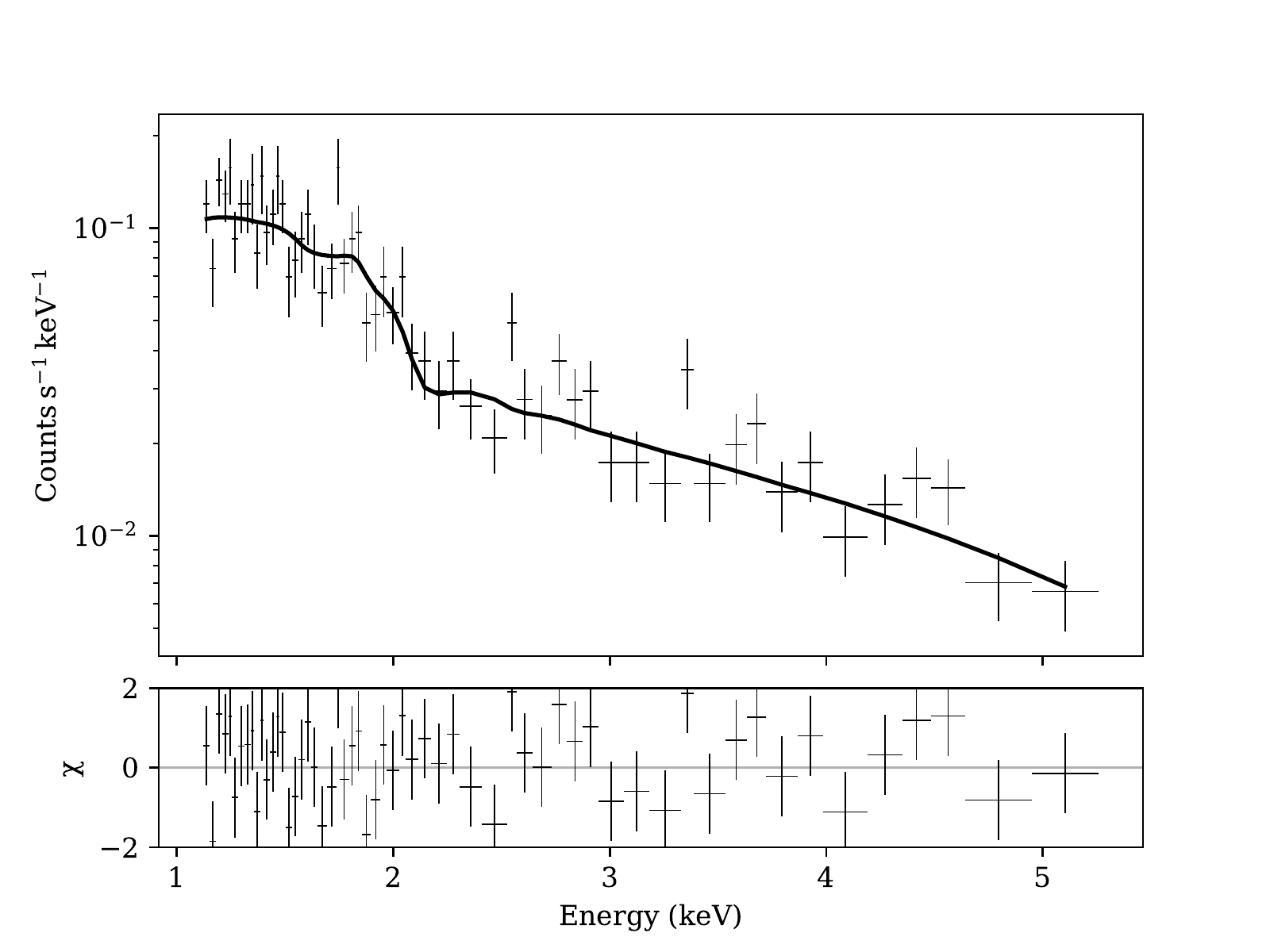}{0.3\textwidth}{(e)}
          \fig{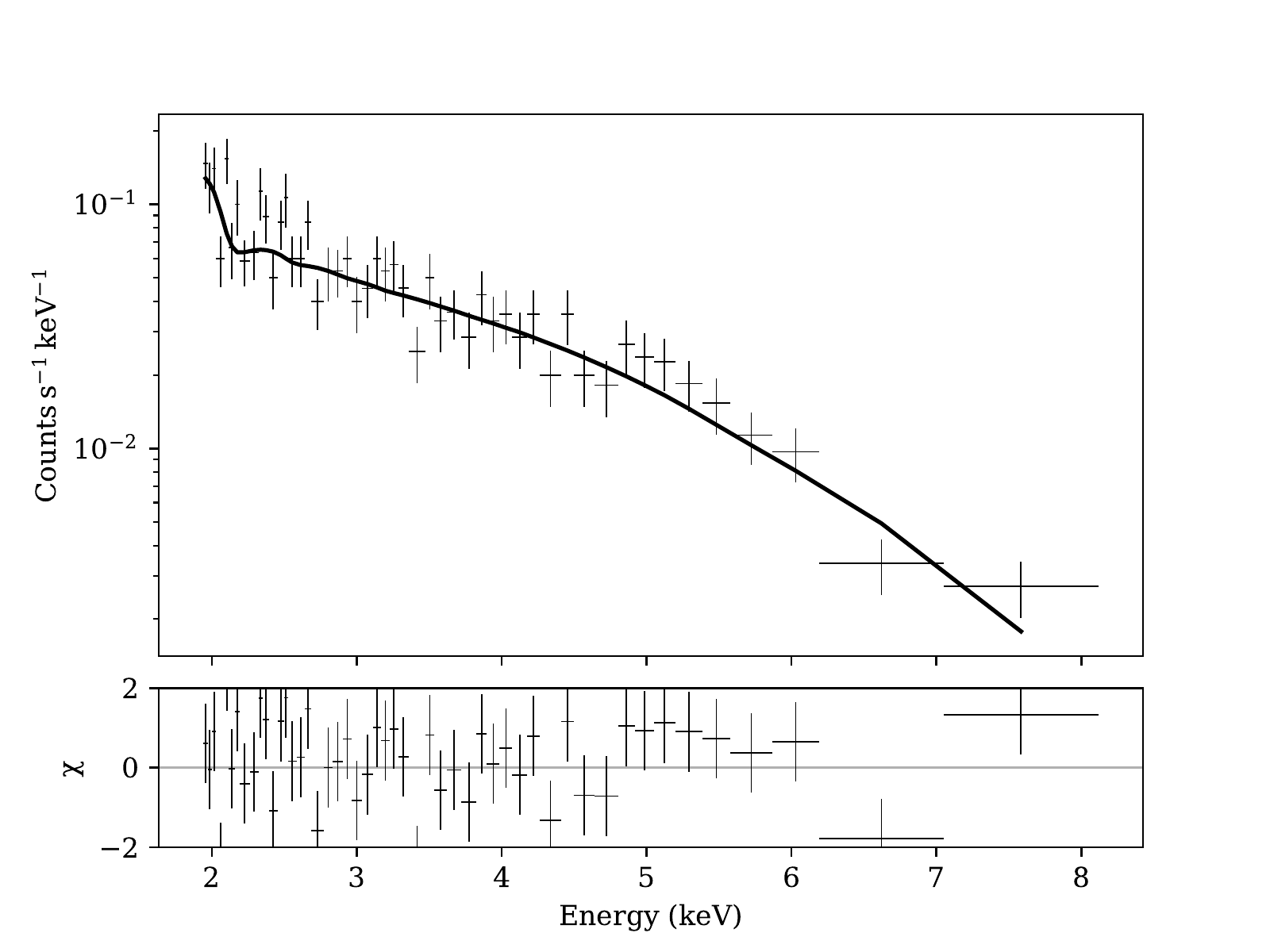}{0.3\textwidth}{(f)}
          }
\gridline{\fig{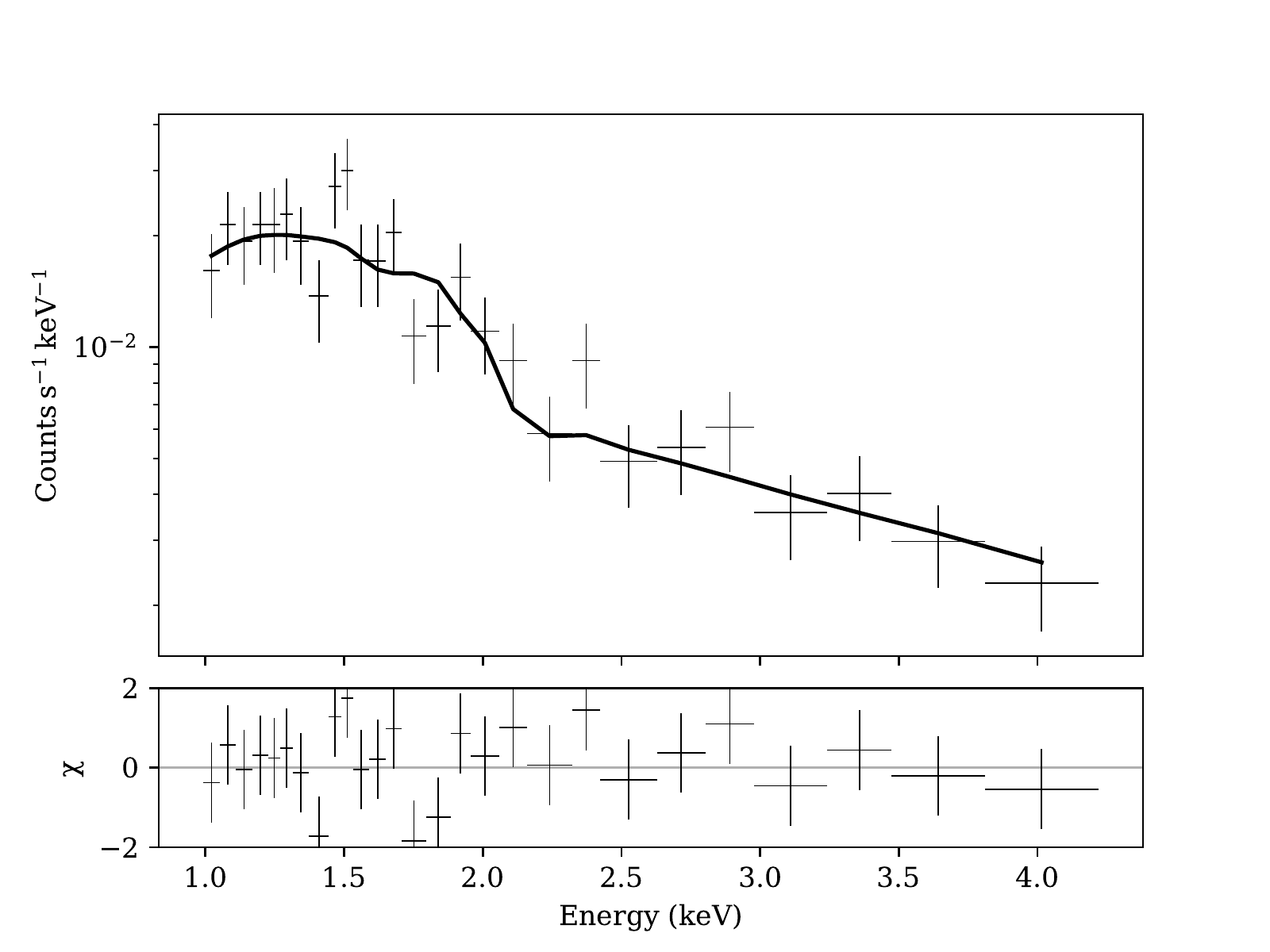}{0.3\textwidth}{(g)}}
\caption{The \chandra{ }X-ray spectra and best fit models of the seven quasars in the sample. The quasars are (a) \nineteenfivetwofive{}, (b) \nineteenfivetwoeight{}, (c) \RBS{}, (d) \PG{}, (e) \FBQS{}, (f) \Mrk{}, and (g) \fourC{}. The spectra are plotted over the rest frame 2-10 keV band appropriate for each source's redshift. Error bars show 1$\sigma$ confidence intervals.}
\label{fig:allxspecs}
\end{figure*}

We include X-ray spectra for the full sample here, including the best fit models and resulting residuals (Figure~\ref{fig:allxspecs}).

\end{document}